\documentclass[
 reprint,
showpacs,preprintnumbers,
 amsmath,amssymb,
 aps,
prb,
url
]{revtex4-1}
\usepackage{color}
\usepackage{mathastext}
\usepackage{graphicx}
\usepackage{dcolumn}
\usepackage{bm}

\begin{document}
\bibliographystyle{apsrev4-1}

\preprint{APS/123-QED}

\title{Tuning the magnetic and structural phase transitions of PrFeAsO via Fe/Ru spin dilution}

\author{Yuen Yiu$^a$, Pietro Bonf\`{a}$^b$, Samuele Sanna$^{c}$, Roberto De Renzi$^b$, Pietro Carretta$^c$, Michael A. McGuire$^d$, Ashfia Huq$^e$, and Stephen E. Nagler$^{a,f,g}$
}
\affiliation{
$^a$Department of Physics and Astronomy, University of Tennessee;
$^b$Department of Physics and Earth Sciences, University of Parma;
$^c$Department of Physics, University of Pavia;
$^d$Materials Science and Technology Division, Oak Ridge National Laboratory;
$^e$Chemical and Engineering Materials Division, Neutron Sciences Directorate,  Oak Ridge National Laboratory;
$^f$Quantum Condensed Matter Division, Oak Ridge National Laboratory.
$^g$CIRE, University of Tennessee.
}%

\date{\today}

\begin{abstract}
Neutron diffraction and muon spin relaxation measurements are used to obtain a detailed phase diagram of PrFe$_{1-x}$Ru$_{x}$AsO. The isoelectronic substitution of Ru for Fe acts effectively as spin dilution, suppressing both the structural and magnetic phase transitions.  The temperature, $T_S$, of the tetragonal-orthorhombic structural phase transition decreases gradually as a function of x.  Slightly below $T_S$ coherent precessions of the muon spin are observed corresponding to static magnetism, possibly reflecting a significant magneto-elastic coupling in the FeAs layers.  Short range order in both the Fe and Pr moments persists for higher levels of x.  The static magnetic moments disappear at a concentration coincident with that expected for percolation of the J$_1$ - J$_2$ square lattice model. 


\end{abstract}

\pacs{28.20.Cz, 65.40.De, 76.75.+i, 74.20.Mn}
\maketitle


\section{\label{sec:level1}Introduction}

The precise role of magnetism and its coupling to the lattice is a central problem in the physics of unconventional iron based superconductors and related materials \cite{Lumsdenreview, Johnston2010, Tranquada2014}. In general, the undoped parent compounds of the 1111 family iron-pnictide superconductors are tetragonal paramagnets at high temperatures.  Upon cooling they display a tetragonal-orthorhombic structural transition at $T_S$, followed or accompanied by a spin density wave (SDW) transition at $T_{SDW}$. \cite{Johnston2010, Stewart2011, Johrendt2011}.  Superconductivity can usually be induced by suppressing these transitions and inevitably results when this is done by using dopants that introduce charge carriers.  The use of isovalent dopants, for example the substitution of Ru for Fe, allows for investigations of the physics without the complications induced by changing the electron count. In the 122 family compound BaFe$_{2-x}$Ru$_x$As$_2$ the suppression of the structural and magnetic transitions via Ru substitution indeed results in a superconducting ground state, but at a much larger Ru content than has been observed with non-isovalent dopants \cite{Thaler2010, Kim2011}.  Investigations of the 1111 compounds including PrFe$_{1-x}$Ru$_{x}$AsO and LaFe$_{1-x}$Ru$_{x}$AsO also showed that Ru/Fe substitution suppresses the structural and magnetic phase transitions but in contrast to the 122s, there is no observation of a superconducting ground state for any concentration of Ru.\cite{mcguire09a, Bonfa2012, yiu12, Martinelli2013}.

Arguably, when Ru is substituted for Fe in the 1111 compounds, the main effect on the magnetism can be understood by considering the substitution  as simply equivalent to spin dilution. This is consistent with local density approximation calculations on LaFe$_{1-x}$Ru$_{x}$AsO \cite{tropeano10}, which illustrates that Ru atoms do not show any tendency to sustain a magnetic moment regardless of their concentration.  This is also compatible with previous experimental data on PrFe$_{1-x}$Ru$_{x}$AsO \cite{mcguire09a}.  In this paper, we present a systematic study of the evolution of the magnetic and structural transitions in the isovalently doped PrFe$_{1-x}$Ru$_{x}$AsO system.  The neutron diffraction measurements of Ref.\onlinecite{yiu12} have been extended and complemented by new muon spin relaxation measurements. The previous neutron work \cite{yiu12} showed no evidence for the structural transition in PrFe$_{1-x}$Ru$_{x}$AsO above $x = 0.4$ as determined by Rietveld refinements.   The magnetic transition in the FeAs layers was not detected beyond $x = 0.1$.  The x = 0.1 sample was previously measured using elastic scattering at the HB1A triple axis spectrometer\cite{yiu12}, and was not sensitive to ordered moment sizes less than $0.02 \mu_B$.  The other neutron diffraction measurements had significantly lower sensitivities. The sensitivity of $\mu$SR ($\approx 0.001 \mu_B$) is therefore more than an order of magnitude better than the neutron measurements, enabling a more complete determination of the phase diagram. Remarkably, it is observed that all signatures of magnetic order disappear at the percolation concentration of the $J_1-J_2$ square-lattice model.  We also note that the negative thermal expansion (NTE) reported earlier in PrFe$_{1-x}$Ru$_{x}$AsO \cite{Kimber08,yiu12} persists across the entire Ru doping range even for pure PrRuAsO.

This paper is organized as follows: Sample synthesis is described in section II, bulk characterization and neutron diffraction results in section III, and $\mu$SR results in section IV.  Discussion and conclusions follow in sections V and VI respectively.

\section{\label{sec:level1}Sample Synthesis}

Methods reported earlier \cite{mcguire09a, mcguire12} were used to synthesize the samples. PrFe$_{1-x}$Ru$_{x}$AsO samples were made from powders of PrAs, Fe$_2$O$_3$, RuO$_2$, Fe and Ru.  The starting materials were crushed and mixed inside a He glovebox, then pressed into a 1/2"  diameter pellets ($\sim$ 2g each) and placed in covered alumina crucibles inside silica tubes.  The tube was evacuated, backfilled with ultra-high-purity Ar and flame sealed.  Each individual sample was heated at 1200$^\circ$C for $12 - 36$h several times, and was thoroughly ground and pressed into pellets between the heating cycles.

\section{\label{sec:level1} Bulk characterization and Neutron diffraction results}

Heat capacity and dc magnetic susceptibility measurements were performed using the MPMS SQUID and PPMS system by Quantum Design.  Fig.1(a) shows the reciprocal magnetic susceptibility temperature dependence for x = 1, i.e. PrRuAsO, with no indication of superconductivity down to 2K.  The Curie-Weiss law describes the data well down to 14K, coinciding with the Ne\'{e}l temperature for Pr ordering in PrFeAsO.  Data points for $T > 50K$ were fitted to the Curie-Weiss law, with the resultant fit intersecting the temperature axis at T$_{CW}$ = -33(5)K with a Curie constant of 1.4(1), close to the expected value of 1.6 for Pr$^{3+}$.  Fig.1(b) shows the field dependence of the magnetization at T = 2K, with no sign of saturation up to 6T.  Fig.1(c) shows the temperature dependence of the heat capacity. A broad hump was observed around 14K, 
the same temperature where the anomaly in the $1/\chi$ due to the Pr ordering is detected (Fig.1(a)).

\begin{figure}
\includegraphics[width=88mm]{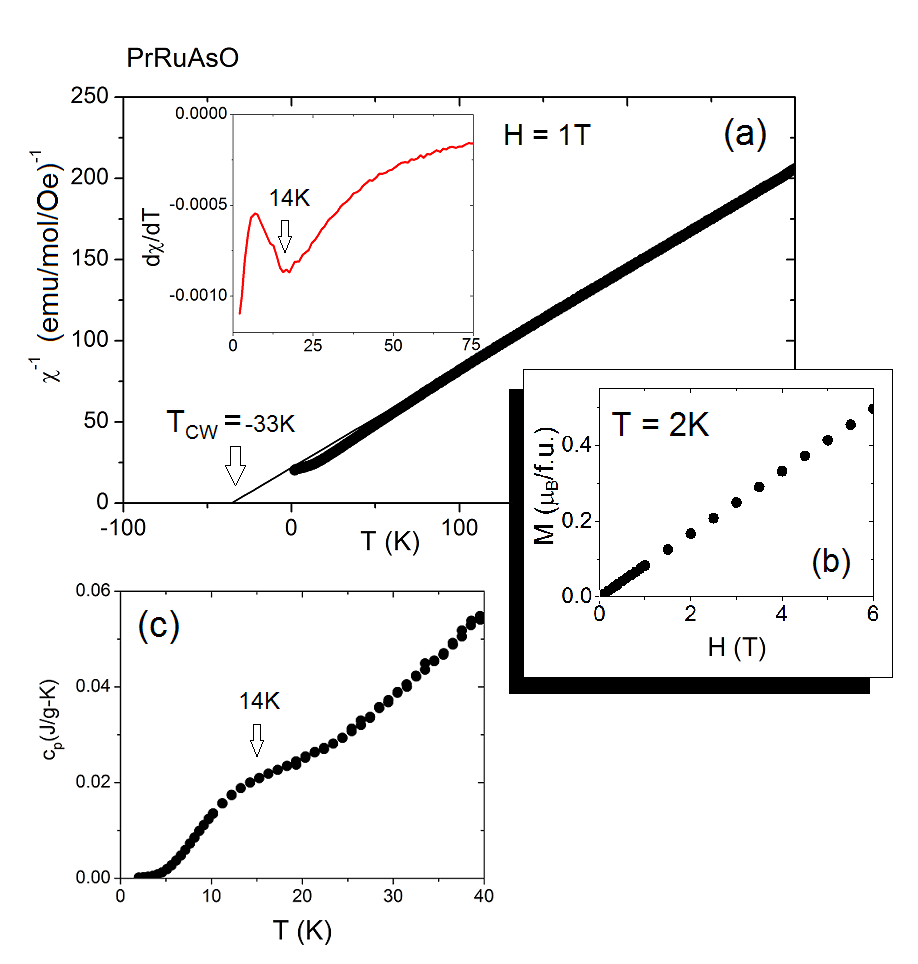}
\caption {(a):  Temperature dependences of the reciprocal susceptibility for PrRuAsO.  The anomaly around 14K is possibly related to magnetic ordering of the Pr sublattice.  The straight line is the Curie-Weiss law fitted to data above 50K.  (a, inset): d$\chi$/dT vs T, better illustrating the anomaly at 14K.  (b): Field dependence of the magnetic susceptibility at T = 2K, with no sign of moment saturation up to 6T; (c):  Temperature dependence of heat capacity.  Similar to that from the $x = 0.1 - 0.75$ samples, the sharp peak in PrRuAsO at T$_{N,Pr}$ = 14K is suppressed, but a broad hump remains \cite{yiu12}.}
\end{figure}

Neutron powder diffraction was performed using POWGEN at the Spallation Neutron Source of Oak Ridge National Laboratory.  Rietveld refinement of the data confirmed that PrRuAsO is isostructural to PrFeAsO at room temperature, and remains in the tetragonal \textit{P4/nmm} structure down to the base temperature of 10K, similar to other PrFe$_{1-x}$Ru$_{x}$AsO samples with $x \ge 0.33$\cite{yiu12}.  Some of the figures below include data previously reported\cite{yiu12}.  In Ref.\onlinecite{yiu12} the neutron diffraction was analyzed via Rietveld refinement, and the transition temperature was determined by whether or not the quality of fit was better for the orthorhombic or tetragonal structure.  For x=0.4 the difference in quality of fit was undetectable down to base temperature, and it was concluded that the structure was tetragonal.  As described in the next paragraph, here we have re-analyzed that data using a different criterion for identifying the transition.

Fig.2 shows the temperature dependence of the orthorhombicity, defined as $(a-b)/(a+b)$. At high temperature all of the samples are tetragonal and the orthorhombicity is zero by definition. The orthorhombicity values plotted in Fig.2 were determined as follows: lattice parameters a and b were extracted by imposing an orthorhombic structure on the Rietveld refinement over the entire temperature range for all samples.  For each doping concentration the fitted value $(a-b)/(a+b)$ determined by the refinement to an orthorhombic structure at T = 200K was subtracted from the corresponding values at other temperatures.   This analytical method is useful for detecting structural transitions that are too subtle to be observed via the splitting or broadening of a single nuclear Bragg peak.  The data shows clear evidence for the structural transition temperature $T_S$ in samples up to $x = 0.4$. (Here $T_S$ is defined operationally as the temperature at which the orthorhombiciy reaches 1/2 of the asymptotic low temperature value.)  For samples with $x \ge 0.5$, no deviation from zero orthorhombicity can be detected at any temperature.

\begin{figure}
\includegraphics[width=77mm]{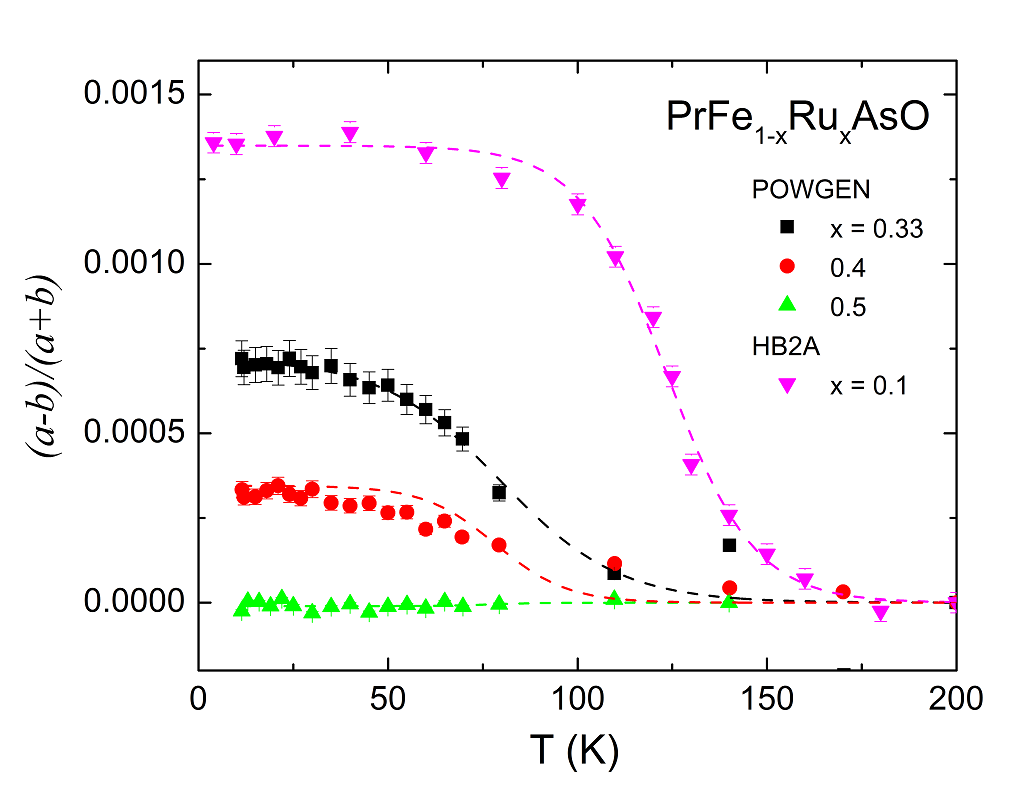}
\caption{Temperature dependence of the orthorhombicity for PrRu$_{1-x}$Ru$_x$AsO as discussed in the text. The lines are guides to the eye.}
\end{figure}

Fig. 3(a) shows the doping dependence of lattice parameters a, b and c (including some data previously published in Ref.\onlinecite{yiu12}).   As Fe is substituted by Ru, the in-plane (a,b)-axis elongates, and the out-of-plane c-axis shrinks \cite{mcguire09a}.  The difference of the lattice parameters between x = 0 and x = 1 is of the order of a few \%.  Fig.3(b) shows the temperature dependence of a and c for PrRuAsO. As reported previously \cite{yiu12, Kimber08}, for $x \le 0.75$ PrFe$_{1-x}$Ru$_x$AsO exhibits NTE in the c-axis for temperatures below approximately 50K. The NTE is also observed clearly in stoichiometric PrRuAsO. The magnitude of NTE in the c-axis is about 0.02\% relative to the minimum at 50K.  The a-axis shrinks more than that predicted by the Debye-Gr\"{u}neisen model, and compensates somewhat for the NTE in the c-axis, resulting in a smaller NTE as determined by the unit cell volume shown in Fig. 3(c).  This compensating behavior can be explained by considering that an expansion in the a-b plane forces the unit cell to shorten along c in order to satisfy Fe/Ru-As bonding requirements \cite{mcguire09a}.   The opposite signs of the x-dependence of the a-b and c lattice parameters can similarly be understood.  The in-plane expansion as a function of x has been attributed to the substitution of larger Ru atoms for Fe atoms, which stretches along the a-b plane \cite{mcguire09a}.  

\begin{figure}
\includegraphics[width=77mm]{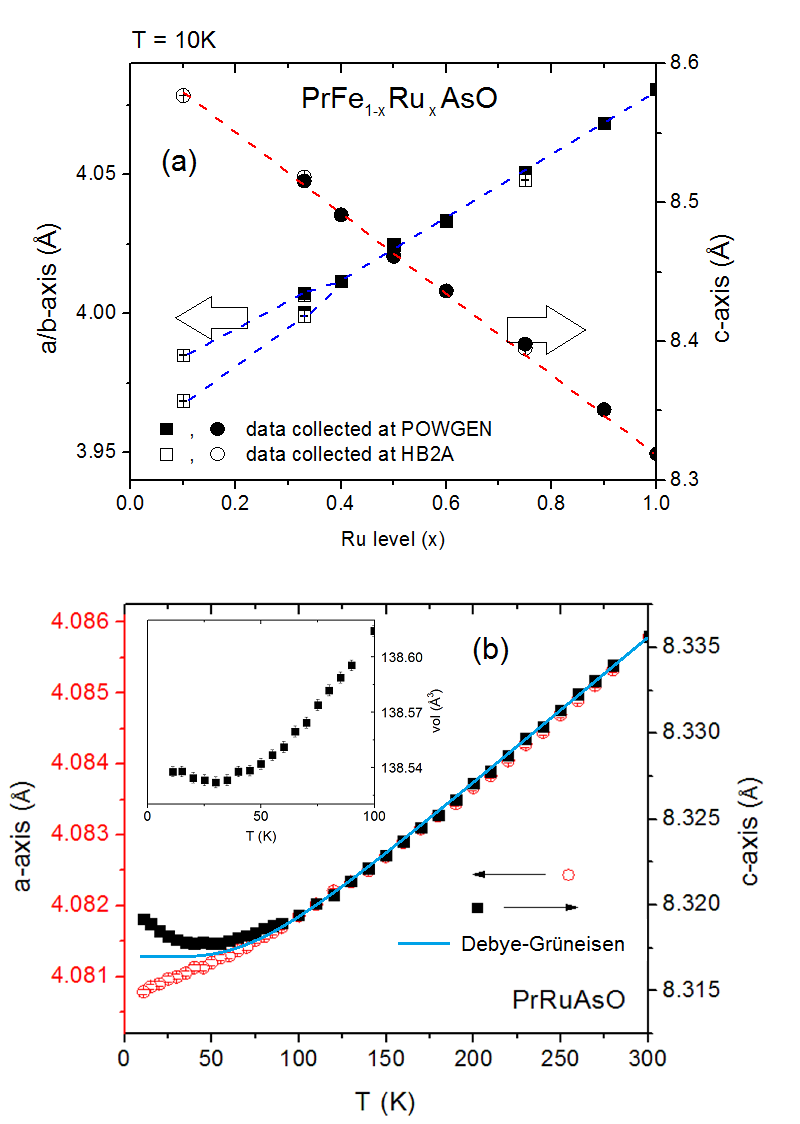}
\caption{(a): Doping dependence of lattice parameters a, b and c. Data for $x \neq 1$ is from Ref.\onlinecite{yiu12}; (b):  T-dependence of lattice parameter a and c  for PrRuAsO.  The solid (blue) line shows Debye-Gr\"{u}neisen fit (see Ref. \onlinecite{yiu12}) for $T \ge 100$K.  The inset shows the T-dependence of the cell volume.  The effect of NTE in the c-axis is somewhat compensated by the opposing behavior of a-axis.}
\end{figure}

\section{\label{sec:level1} $\mu$SR}

\begin{figure}
\includegraphics[width=77mm]{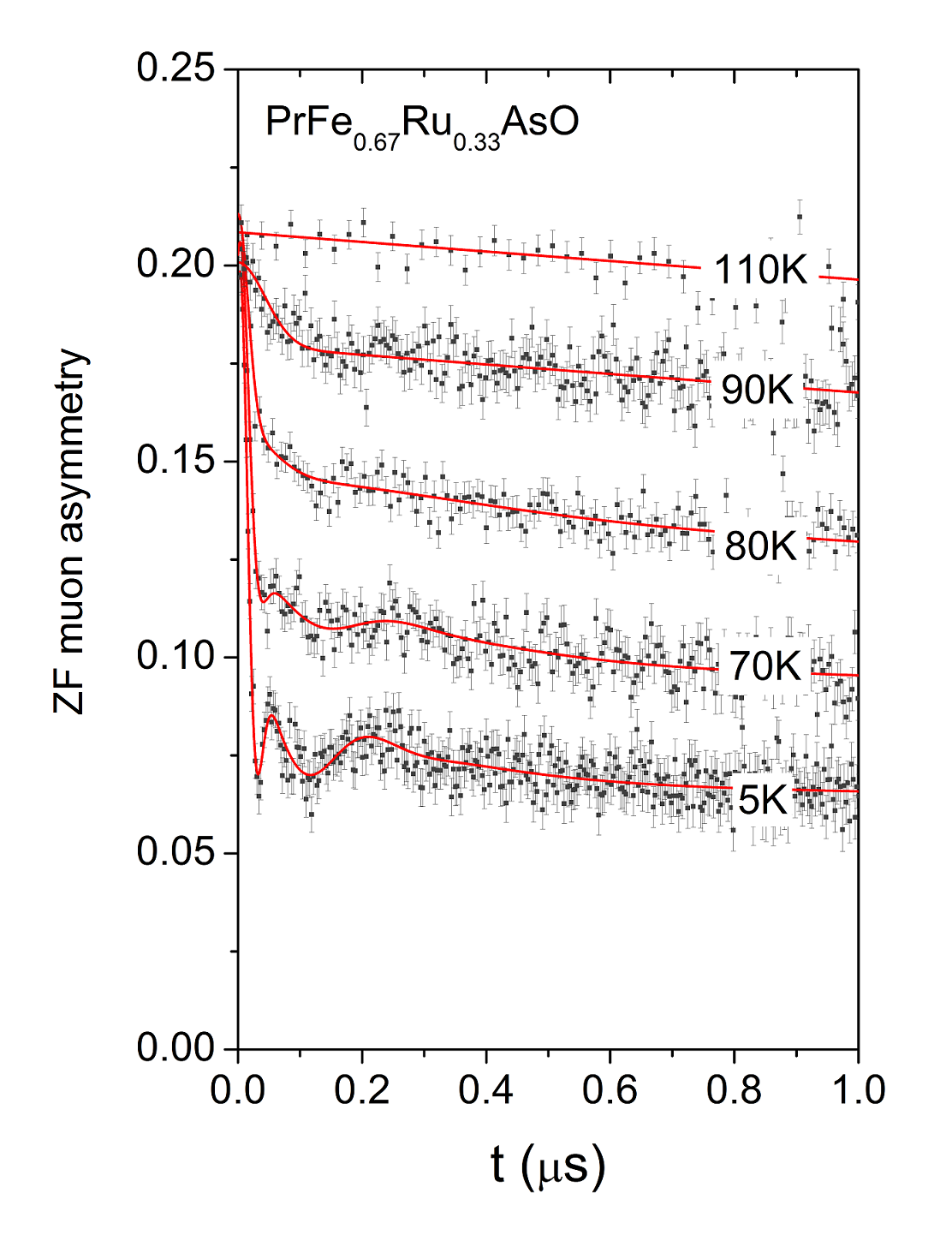}
\caption{Time dependence of ZF-$\mu$SR asymmetry for x = 0.33 between 110K and 5K, the lines represent the best fit using Eq.\ref{Eq:ZFmusrfit}.}
\end{figure}

\begin{figure}
\includegraphics[width=66mm]{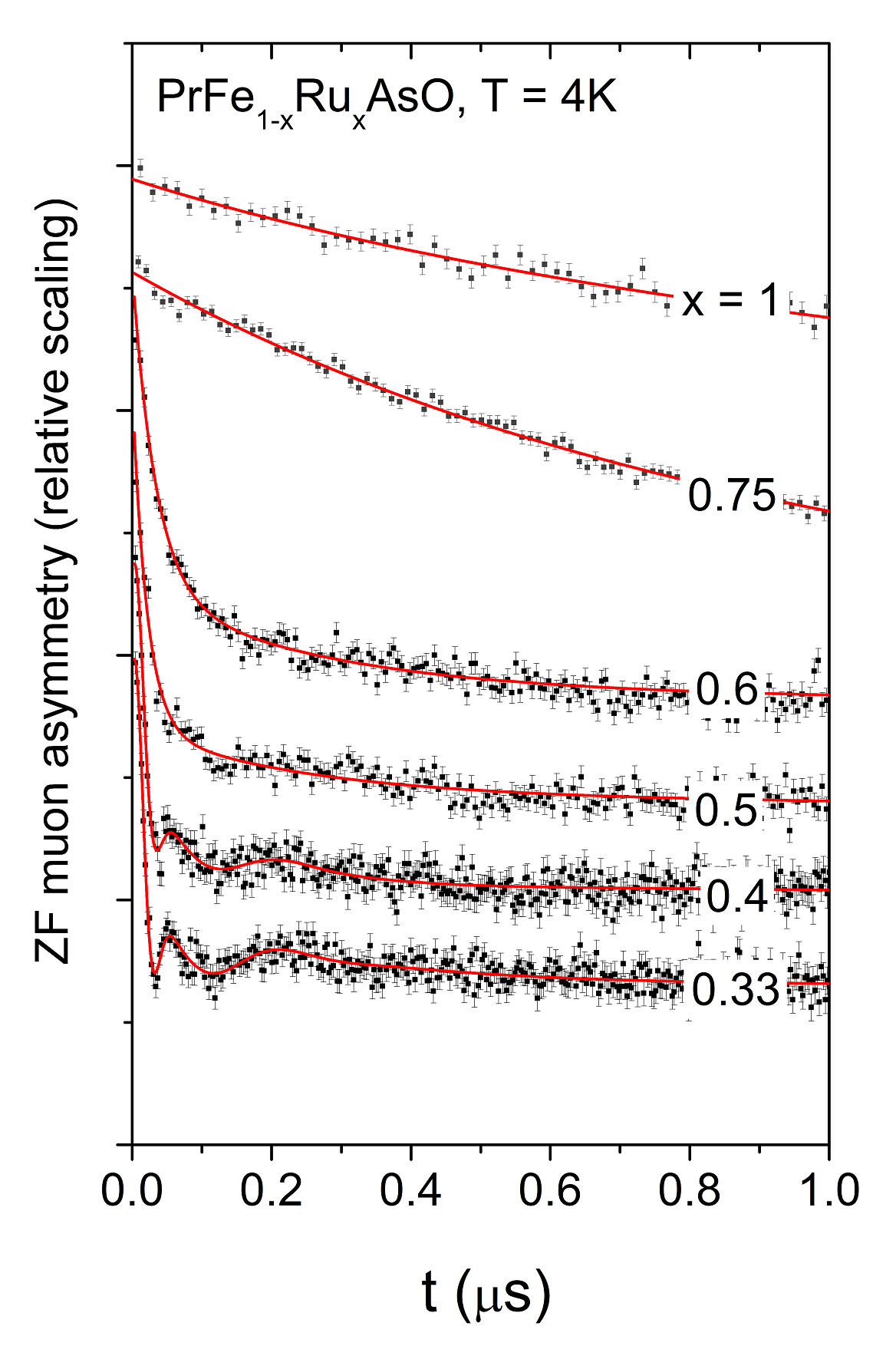}
\caption{Time dependence of ZF-$\mu$SR asymmetry for x = 0.33, 0.4, 0.5, 0.6, 0.75 and 1 at T = 4K. For graphing purposes, each composition is shifted along the vertical axes by an arbitrary constant.}
\end{figure}

Zero field (ZF) and longitudinal field (LF) $\mu$SR experiments were performed on the GPS spectrometer at the Laboratory for Muon Spin Spectroscopy of Paul Sherrer Institut.  Here the findings are presented  in two parts, one for the SDW ordering involving FeAs layers, and the other for the magnetic ordering of the Pr moment.

\subsection{\label{sec:SDWordering} SDW ordering}

The following methods were used to interpret the data.  For the ZF-data, the time dependence of the spin-polarization function for a positive, 100\% spin-polarized muon in a magnetic sample can be described as:

\begin{eqnarray}\label{Eq:ZFmusrfit}
    \frac{A_{ZF}(t)}{A_{0}} & = & \left(1 - V_{m}\right)
    e^{-\frac{\sigma_{N}^{2}
    t^{2}}{2}} {} \nonumber\\ & + & \sum_{i = 1}^{N}
    f_{i} \left[w_{i}^{\perp} F_{i}(t) e^{-\frac{\sigma_{i}^{2} t^{2}}{2}}+
    w_{i}^{\parallel}    e^{-\lambda_{i} t}\right]\,\,\,
\end{eqnarray}

$A_{ZF}$ is the asymmetry of the muon decay and $A_0$ is the initial muon asymmetry (i.e. t = 0). $V_m$ represents the fraction of muons probing a static local field \textbf{B}$_i$, i.e. the sample's magnetic volume fraction.  The index \emph{i} represents each of N crystallographically-inequivalent muon stopping sites and each stopping site is characterized by a stopping probability $f_i$, with $\sum_{i = 1}^{N} f_{i} = V_m$.  The two terms in the square brackets reflect the  orientation of the internal field with respect to the initial muon spin $S_\mu$ direction: transverse for $\bm{B}_i\perp \bm{S}_\mu$ and longitudinal for $\bm{B}_i\!\parallel\! \bm{S}_\mu$. For powder samples, the ratio of the two terms is related and normalized by $w_{i}^{\perp} = 2/3$ and $w_{i}^{\parallel} = 1/3$.

The longitudinal component ($\parallel$) can be described by a Lorenztian decay function with relaxation rate $\lambda$.  For the transverse component ($\perp$), $F_i(t)$ represents the time dependence, and $\sigma_i$ is the depolarization rate which reflects the second moment of the field distribution $\Delta B_i\!\!\equiv\!\!(\overline{B_i^2}- \overline{B_i}^2)^{1/2} \!\! = \!\! \sigma_i/\gamma_\mu $, where $\gamma_\mu/2\pi = 136$ MHz/T is the muon gyromagnetic ratio.  When the muon goes through a local field $B_i$, for example inside a long range ordered sample, the muon asymmetry displays Larmor oscillations described with $F_i= cos(\gamma_\mu B_i t)$, with $B_i$ proportional to the mean magnetic order parameter $<S(T)>$.  In case of a short range ordered sample, the width of the field distribution at the muon site broadens and as a result, the transverse muon fraction yields to a fast decay rate ($\sigma_i\gtrsim 1/\gamma_\mu B_i$), with overdamped oscillations and $F_i = 1$.  

For the undoped parent PrFeAsO the ZF-$\mu$SR time spectra are well fitted with $N = 2$ and occupancy $f_1 \approx 75\%$ and $f_2 \approx 25\%$, which reflects the presence of two inequivalent muon sites\cite{Maeter2009,Derenzi2012, Prando2013}.  In the x = 0.33 and 0.4 samples, the two frequencies are still detectable, but with $f_2$ reduced in half. The complementary missed amplitude gives rise to overdamped oscillations and can be easily fitted as a third additional non-oscillating component i = 3, with $f_3 \approx f_2$, $F_3 = 1$ and $\sigma_3 \sim 5 \mu s^{-1}$. This change might be simply due to the increase of disorder by Ru.

These three components provide a good fit of the time evolution of the ZF muon asymmetry, as seen in Fig.4, which shows data for $x = 0.33$ at different temperatures, fitted with eq.\ref{Eq:ZFmusrfit}.  At higher temperatures the oscillations become overdamped and the transverse amplitude $(\propto V_m)$ reduces and vanishes at $T \geq 100$K.  Fig.5 displays the low temperature ZF-$\mu$SR time spectra for all our samples.  For x = 0.5 and 0.6 the coherent oscillations are absent and the fit uses only 2 components but with $F_1 = F_2 = 1$, suggesting that the increase of Ru/Fe substitution induces field inhomogeneity.  Accordingly, the presence of the magnetic phase is reflected by the sizeable decay rate detected corresponding to $\Delta B_1 \approx 40$ mT and $\Delta B_2 \approx 5$ mT.  The same behavior has also been reported in Ru/Fe substituted LaFeAsO \cite{Bonfa2012} at a similar doping level.

Fig.5 shows the components with fast decay rates in samples from x = 0 up to 0.6. The lack of a fast decay component in the x = 0.75 and 1 samples indicates that Fe moments do not order in those samples.  However, for T below $\approx 14$K, the fit of the LF muon asymmetry requires two non-oscillating amplitudes with distinct relaxation rates, as shown in Fig.6(a) and 6(b).  This behavior can be attributed to activities in the Pr sublattice around $T_{Pr} \approx 14$K, where there is noticeable features in both the susceptibility and specific heat measurements (Fig.1 and Ref. \onlinecite{yiu12}).  We will discuss this point again later in section IV-B.

\begin{figure}
\includegraphics[width=70mm]{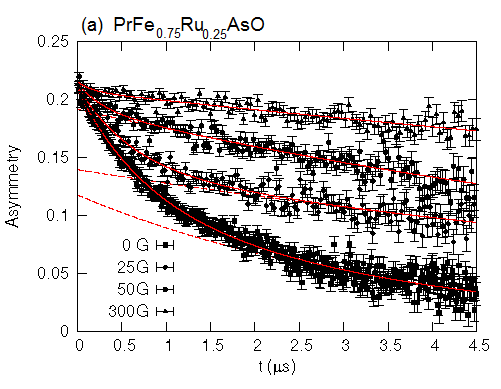}
\includegraphics[width=70mm]{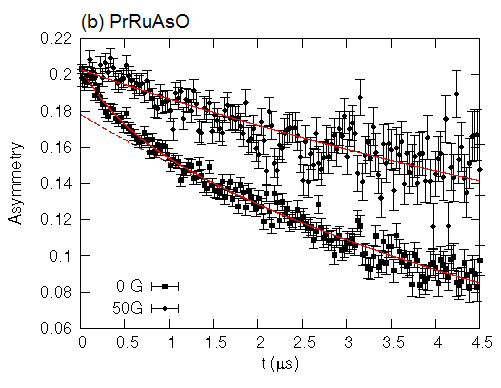}
\caption{LF-$\mu$SR time spectra for (a):  PrFe$_{0.25}$Ru$_{0.75}$AsO and (b):  PrRuAsO, for different external fields $H\!\parallel\! \bm{S}_\mu$. The lines are fits to the two non-oscillating amplitudes with distinct relaxation rates.}
\end{figure}

\begin{figure}
\includegraphics[width=75mm]{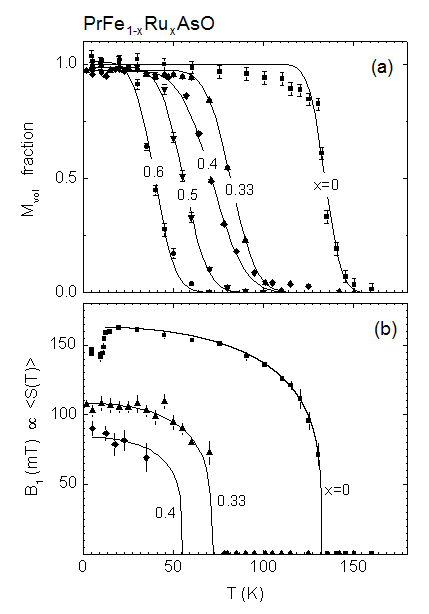}
\caption{ (a): Magnetic volume fraction as a function of temperature for x = 0.33, 0.4, 0.5 and 0.6, with the main contribution from SDW ordering. The lines are guides for the eye fits; (b): Temperature evolution of the internal field at the muon site, $B_1$, proportional to the mean magnetic order parameter $<S>$. The lines are the best fit to a phenomenological mean field-like function as described in the text. }
\end{figure}

To summarize, the temperature dependence of the magnetic volume fraction for all samples is shown in Fig.7.  We are able to detect ordering in the FeAs layer up to $x = 0.6$.  For the construction of a phase diagram later in this paper, we define $T_{SDW}$ as the temperature at which the magnetic volume fraction is 80\%. For $x \leq 0.40$, the magnetic transition temperatures, $T_{SDW}$, can be directly determined from the evolution of the mean magnetic order parameter $<S(T)>\propto B_i (T)$ as a function of temperature, shown in Fig.7(b). The magnetic order parameter $<S(T)>$ has been successfully fit to the phenomenological function $<S(T)>=S(0) [1-(T/T_{SDW})^{2.4}]^{0.24}$, which is found to hold generally for REFeAsO compounds \cite{Maeter2009}. The values of $T_{SDW}$ determined using the two criteria are consistent to within about 2K. 

\subsection{\label{sec:Prordering} Pr magnetic ordering}

Previous neutron diffraction measurements did not detect long range ordering (LRO) of Pr in samples with $x \ge 0.1$ \cite{yiu12}.  However, the $\mu SR$ data shows that the muon relaxation rate increases below $T \sim 14$ K in PrFe$_{0.25}$Ru$_{0.75}$AsO (see Fig.6(a)), hinting at a possible short range ordered (SRO) state involving the Pr sublattice.  Fig.6(b) shows the LF-$\mu SR$ spectra for PrRuAsO.  For x = 0.75, the muon relaxation function ($A_{ZF}$) in zero magnetic field consists of two separate components: $A_f e^{-\lambda_f t} + A_s e^{-\lambda_s t}$, where $\lambda_f = 1.4 \mu$s$^{-1}$ represents the fast decay rate and $\lambda_s = 0.25\mu$s$^{-1}$ represents the slower one.

This empirical fitting function mimics the trend expected for a quasi-static Kubo-Toyabe relaxation\cite{Kubodyn} with a Lorentzian distribution of internal fields having HWHM $\Delta B = \lambda _f / \gamma_\mu \sim 2$~mT and characterized by a slow dynamics with correlation time $\tau \sim 1/ \lambda_s$.  In the static case ($\tau \rightarrow \infty$) $A_f = 2A_s$, but $A_s$ is expected to grow as dynamics sets in.

The onset of longitudinal fields decreases the fast decay rate and increases the slow decay rate in both samples.  For x = 0.75, a field of $H = 300$G completely suppresses the faster (static) component of the muon relaxation and only leaves the slower dynamical component.  This suggests that the ordering is quasi static. The observed values of $\Delta B$ are consistent with the dipolar field from Pr moments ($\sim 3.6 \mu_B$).  Fig.6(b) shows the same for $x = 1$. In this case a longitudinal field $H = 50$G completely suppresses the faster (static) component of the zero field relaxation and the residual dynamics is a little bit faster.

\section{Discussion}

\begin{figure}
\includegraphics[width=82mm]{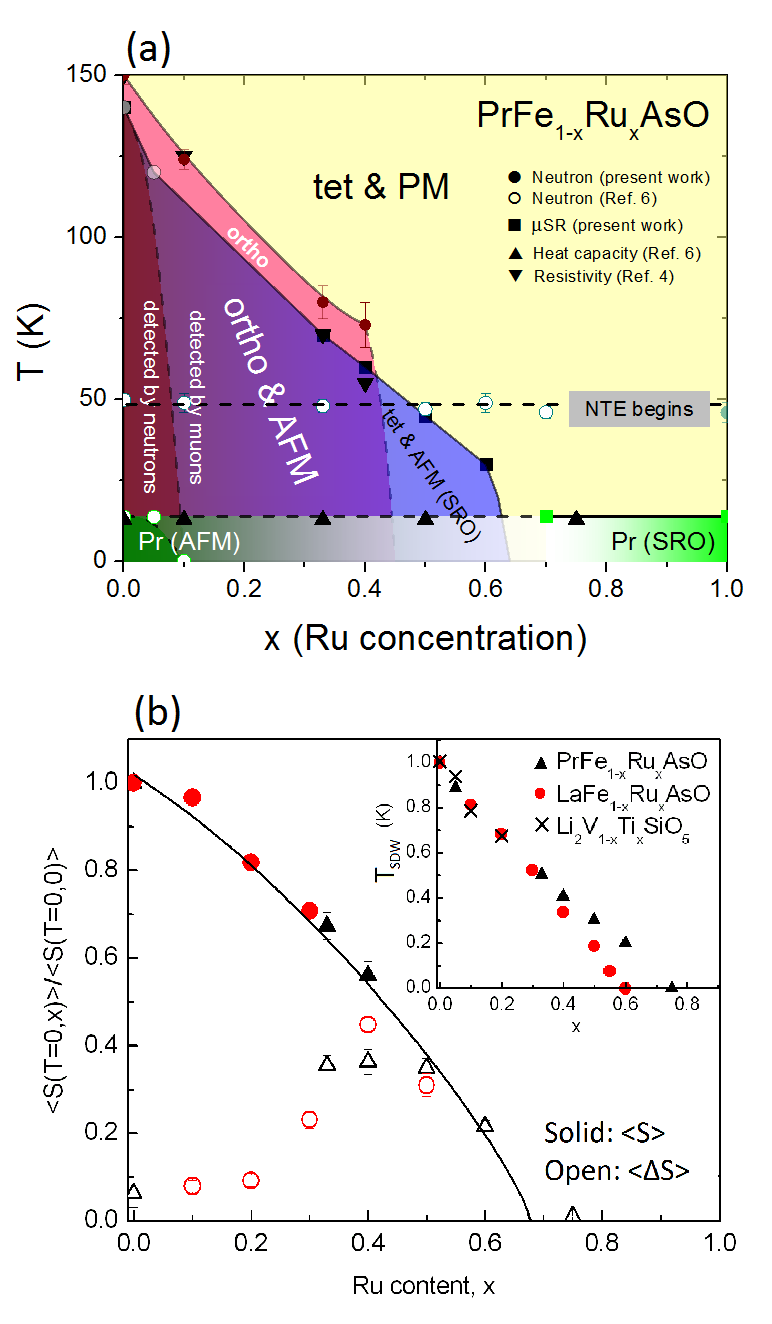}
\caption{(a): A phase diagram for PrFe$_{1-x}$Ru$_{x}$AsO constructed using data presented in this paper and also data from Ref. \onlinecite{yiu12}.  (tet: Tetragonal; ortho: Orthorhombic; AFM: Antiferromagnetic; PM: Paramagnetic; SRO: Short range order; NTE: Negative thermal expansion) The location of the structural transition is defined by the orthorhombicity analysis discussed in the text; (b) The doping dependence of the staggered magnetization at zero temperature $S(0,x)$ (solid symbols) and of its distribution width $\Delta S(0,x)$ (open symbols) in PrFe$_{1-x}$Ru$_{x}$AsO (triangles) and LaFe$_{1-x}$Ru$_{x}$AsO (circles, from Ref.\onlinecite{Bonfa2012}). The data are normalized to the value of the undoped member $S(0,0)$; (b,inset): Doping dependence of magnetic transition temperatures, $T_{SDW}$, normalized to the undoped value for PrFe$_{1-x}$Ru$_{x}$AsO, LaFe$_{1-x}$Ru$_{x}$AsO and Li$_2$V$_{1-x}$Ti$_x$SiO$_5$.}
\end{figure}

Fig.8(a) displays the T vs x phase diagram for PrFe$_{1-x}$Ru$_x$AsO as determined by neutron diffraction and $\mu$SR, with some additional points determined from anomalies detected in heat capacity and resistivity measurements.  Using the orthorhombicity criterion described above, the tetragonal to orthorhombic structural transition is detected by neutron scattering for $x$ up to 0.4.  Long range magnetic order as detected by neutron diffraction is observed only for $ x \le 0.1$.  Conversely, from the $\mu$SR data one detects signs of static moments in the FeAs layers to approximately $x=0.6$, and evidence for Pr moments up to x = 1.  To reconcile these observations it must be noted that the neutron measurements of magnetic Bragg peaks are sensitive to spatially dependent long range order.  Conversely, muons provide a sensitive probe of local magnetic fields, and therefore may detect local fields associated with static short range order, i.e. with magnetic moments that fluctuate slowly (correlation times less than 1 $\mu$s) and with a short magnetic coherence length (even less than 10 lattice spacings) \cite{Routier}.  Such SRO does not contribute to the magnetic Bragg peaks detected via neutron diffraction.  

The muon data for samples with $0.4 \le x \le 0.6$ exhibits overdamped oscillations and the $\mu$SR asymmetry displays a component with a fast decay rate ($\sigma_1 \sim 50\mu s^{-1}$), indicating that the system is still magnetically ordered but the muon spin precessions have become incoherent. Conventionally, this implies that the correlation length of the Fe ordered domains has become shorter than about 10 unit cells \cite{Routier}. This suggests a transition from LRO to SRO as detected by muons.  This is indicated by a dashed vertical line in Fig.8(a) around x = 0.4, which interestingly coincides with the suppression of the structural transition as detected by neutrons.  The fact that the progressive reduction of $T_{SDW}$ is closely accompanied by the reduction of $T_S$ hints at significant magneto-elastic coupling in the FeAs layers.  

The persistence of static moments as observed by $\mu$SR despite the absence of LRO implied by the neutron measurements may provide a clue to the reason for the failure of superconductivity to appear in PrFe$_{1-x}$Ru$_{x}$AsO.  

Notwithstanding the fact that the coexistence of magnetic order and superconductivity is possible in Fe pnictide compounds, superconductivity almost always emerges in these materials when both the structural and magnetic transitions are suppressed.  However there is evidence that a necessary condition for the emergence of superconductivity is the persistence of magnetic fluctuations.  Some indirect evidence for the latter statement is the complete absence of magnetic fluctuations in the collapsed tetragonal phase observed in 122 family materials\cite{Soh2013}. One can speculate that mutually uncorrelated but effectively frozen Fe spin clusters might exist in randomly diluted PrFe$_{1-x}$Ru$_{x}$AsO at low temperatures. As a consequence a static moment might be detected by a local probe such as $\mu$SR, and the absence of significant magnetic fluctuations would impede the emergence of a superconducting ground state.

Fig.8(b) shows the evolution of the magnetic order parameter $<S(T\rightarrow 0)>$ vs x as determined by $\mu$SR. The spin dilution caused by Ru/Fe substitution reduces both the ordering temperature $T_{SDW}$ and the moment size gradually.  The complete suppression of the SDW ordering is determined to be around x = 0.6 which is similar to the concentration inferred from previous resistivity measurements\cite{mcguire09a}.  Perhaps coincidentally, this is very close to the disruption of superconductivity by Ru/Fe substitution in F-optimally doped 1111 \cite{Satomi2010, Sato2010, Sanna2011, Sanna2013}.  This value may be very significant as discussed below.

A proper description of the magnetism in the Fe-pnictides must account for the fact that the systems are itinerant, however despite this many of the main features can be understood in terms of Hamiltonian models related to local spins.  The two dimensional $J_{1}-J_{2}$ model with Heisenberg nearest neighbor ($J_{1}$) and next nearest neighbor ($J_{2}$) interactions on a square lattice\cite{Chandra1990} exhibits a striped phase for $J_2/J_1 \ge 1/2 $.  Moreover, any non-zero coupling to the lattice results in an Ising-nematic transition associated with a rectangular lattice.  It can be argued on the basis of symmetry that spin driven Ising nematic order must be accompanied by both a structural phase transition and  orbital order \cite{Fernandes2014}, and that nematic order may arise from a correlation driven electronic instability.  In any case, the close association of the orthorhombic structural transition and striped antiferromagnetic order in the iron pnictides inspired many applications of the $J_{1}-J_{2}$ model to explain the underlying physics\cite{Fang2008, Fernandes2010, Abrahams2011, Si2008, Xu2008,  Fernandes2012}.  
Caution must be exercised in applying the $J_{1}-J_{2}$ model to iron pnictide systems.  Inelastic neutron scattering experiments\cite{Wysocki2011, Lynn2009, Zhao2009, Diallo2009} and first-principles calculations\cite{Yin2008} found that fitting observed magnetic excitations using the $J_{1}-J_{2}$ Hamiltonian led to parameters that were physically incompatible with the known ordering scenarios and incapable of giving an acceptable explanation of the response functions.  Since then there has been much work\cite {Hu2012, Yu2012, Wysocki2011, Glasbrenner2014} showing that a minimal effective model must also include a biquadratic exchange term $K(S_1\cdot S_2)^2$ and a small interplane coupling $J_c$.

The biquadratic exchange term must exist in the system and also accounts at least partially for the expected effects of itinerancy.  With this Hamiltonian the observed magnetic excitations can be explained with physically reasonable fitted parameters. Within the context of this expanded model, the scenario for magnetic and structural order remain the same as that expected for the $J_{1}-J_{2}$, with some minor renormalizations of the parameters\cite{Hu2012}.  

As the moment size is reduced the significance of the biquadratic term is also diminished and one expects that the $J_{1}-J_{2}$ model can provide an even better description of the system.  In this context, it is very interesting to consider the vanishing of detectable magnetic order near $x=0.6$.  In the simplest scenario for magnetic dilution, magnetic order is expected if the concentration of magnetic ions (here $1-x$) is greater than or equal to the percolation concentration of the lattice.  When the interactions are of the same magnitude, the percolation concentration of the $J_{1}-J_{2}$ model should be essentially the same as the square lattice with nearest neighbor and next nearest neighbor connectivity.  This model leads to a percolation concentration for magnetic ions almost exactly at the value $1-x=0.4$ \cite{malarz05}.  The fact that this coincides with the disappearance of static magnetism in PrFe$_{1-x}$Ru$_x$AsO is a strong indicator that the core physics of the $J_{1}-J_{2}$ model is at play.  We note that the $J_{1}$ only model exhbits percolation at $1-x=0.59$\cite{malarz05}. The inset of Fig.8(b) compares the doping dependence of $T_{SDW}$ in PrFe$_{1-x}$Ru$_x$AsO to two other systems that also cited the $J_1-J_2$ model for magneto-elastic coupling, namely LaFe$_{1-x}$Ru$_x$AsO\cite{Bonfa2012} and Li$_2$V$_{1-x}$Ti$_x$SiO$_5$\cite{Papinutto2005}, the latter being an archetype of the $S=1/2$ $J_1-J_2$ square lattice model.  The close association of the structural transition with the magnetic order is also explained naturally by the Ising-nematic scenario predicted for localized spins in the $J_1-J_2$ model.  If one takes into account for the presence of a magneto-elastic coupling in the system, the structural transition is closely linked to the occurrence of a spin nematic phase at $T_S$ which anticipates the breaking of the rotational symmetry that is associated with the magnetic transition at $T_{SDW}$. 

As discussed earlier, $\mu$SR shows that SRO of the Pr moments in PrFe$_{1-x}$Ru$_{x}$AsO persists up to $x = 1$.  This is consistent with the anomalies observed in the susceptibility and specific heat measurements around $T_{Pr} \sim 14 K$ (see Fig.1 for the PrRuAsO data ).   The ZF-$\mu$SR results indicates a moderately fast depolarization rate in the muon asymmetry around the same temperature.  This implies the presence of a broad field distribution generated by a non-collinear arrangement of the Pr moments.   LF-$\mu$SR spectra shown in Fig.6 suggests that the magnetic phase is mainly quasi static.  Notably, the NTE seen in the c-axis also persists over the entire range of Ru concentrations. Although $T^{Pr}_N$ and $T_{NTE}$ are markedly different, the continuous presence of both throughout the entire doping range leads one to speculate that there is a relation between the ordering of the Pr sublattice and the NTE, and if the latter is driven by magneto-elastic coupling \cite{Kimber08}, the Pr moments are relevant. Indeed, as shown in Fig. 8(b), the doping dependence of the quantities $T_{SDW}$ and $<S(0)>$ in LaFe$_{1-x}$Ru$_{x}$AsO is similar to that seen in PrFe$_{1-x}$Ru$_{x}$AsO, yet the NTE effect has not been observed in LaFe$_{1-x}$Ru$_{x}$AsO \cite{Martinelli2013}.

\section{\label{sec:conclusions}Conclusion}
In summary, we have combined neutron powder diffraction and muon spin relaxation data for the PrFe$_{1-x}$Ru$_{x}$AsO series, completing the study up to $x = 1$.  The substitution of diamagnetic Ru for magnetic Fe generates a spin dilution process which gradually suppresses the ordering in the FeAs layers, in which evidence for static moments persists until around $x = 0.6$, the magnetic percolation threshold expected under a localized $J_1-J_2$ model \cite{Bonfa2012, malarz05}.  The gradual suppression of the magnetic phase is closely followed by the reduction of the structural tetragonal-orthorhombic phase transition temperature.  The lattice distortion and the magnetic ordering are found to be strongly coupled, as predicted for pnictides by many theoretical works \cite{Si2008, Fang2008, Xu2008, Fernandes2010, Abrahams2011, Fernandes2012}.  The persistence of static moments and possible suppression of magnetic fluctuations may be related to the absence of superconductivity in the system.

In addition, we found that both the magnetic ordering of the Pr sublattice and the negative thermal expansion of the c-axis phenomena persist up to $x = 1$.  We speculate that the abnormal thermal expansion behavior can be linked to the magneto-elastic coupling within the Pr sublattice, which survives despite of the disruption of the ordering of Fe moments.

\section{\label{sec:level1}Acknowledgements}
The research reported here utilized the $\mu$SR facilities at the Paul Scherrer Institute, Villigen, Switzerland, and neutron scattering facilities at the  Spallation Neutron Source, Oak Ridge National Laboratory (ORNL), which is sponsored by the Scientific User Facilities Division of the Office of Science, Basic Energy Sciences, US Department of Energy (BES DOE). We are grateful to A. Amato and H. Luetkens of the Swiss Muon Source group for technical support.  We would like to thank A. A.  Aczel and T. J. Williams of ORNL for valuable conversations. A.H. and S.E.N. were supported by the Scientific User Facilities Division of BES DOE. M.A.M. was supported by the  Materials Sciences and Engineering Division of BES DOE. S.S. and P.C. acknowledge the financial support of Fondazione Cariplo (Research Grant n.2011-0266).  R.D.R., P.B., P.C. and S.S. acknowledge partial support of PRIN2012 project 2012X3YFZ2. Y.Y. was supported by the BES DOE, through the EPSCoR, Grant No. DE-FG02-08ER46528.

\bibliography{paper}

\end{document}